\documentclass[floatfix,showpacs,prd,aps,tightenlines,amsmath,amsfonts,amssymb]{revtex4}

\usepackage{amsmath}
\usepackage{graphicx}
\usepackage{psfrag}
\usepackage{color}
\usepackage{dcolumn}% Align table columns on decimal point
\usepackage{bm}% bold math
\usepackage{longtable}
\newcommand{\vect}[1]{\boldsymbol{#1}}

\newcommand{\bea}{\begin{eqnarray}}
\newcommand{\eea}{\end{eqnarray}}

\def\msun{M_{\odot}}
\def\snr{{\rm SNR}}

\begin{document}

\title{Possible golden events for ringdown gravitational waves}

\author{Hiroyuki Nakano, Takahiro Tanaka and Takashi Nakamura}

\affiliation{Department of Physics, Kyoto University, Kyoto 606-8502, Japan}

\begin{abstract}
There is a forbidden region in the parameter space of 
quasinormal modes of black holes in general relativity.
Using both inspiral and ringdown phases of gravitational waves from
binary black holes, we propose two methods to test general relativity.
We also evaluate how our methods will work when 
we apply them to Pop III black-hole binaries with typical masses.
Adopting the simple mean of the  estimated range of the event rate, 
we have the expected rate of 500 ${\rm yr^{-1}}$. 
Then, the rates of events with 
signal-to-noise ratios greater than 20 and greater than 50
are 32 ${\rm yr^{-1}}$ and 2 ${\rm yr^{-1}}$, respectively. 
Therefore, there is a good chance to confirm (or refute) the 
Einstein theory in the strong gravity region by observing 
the expected quasinormal modes.
\end{abstract}

\pacs{04.30.-w,04.25.-g,04.70.-s}

\maketitle

%%%%%%%%%%%%%%%%%%%%%%%%%%%%%%%%%%%%%%%%
\section{Introduction}
%%%%%%%%%%%%%%%%%%%%%%%%%%%%%%%%%%%%%%%%

Black hole (BH) singularities appear unavoidably
in general relativity (GR). However, as a physics law
the allowance of the presence of singularities will not be 
acceptable even though they are hidden behind the event horizon.
Therefore, various possibilities of the singularity avoidance have
been discussed. Some replacement of singularities 
is required as a complete theory which can describe the BH 
evolution inside the horizon. Although it is totally unknown 
how the singularities are to be regularized, 
there are a lot of proposals motivated by the string theory 
and/or the BH information paradox. 
Some of them, such as gravastars~\cite{Mazur:2001fv}, 
fuzzballs (see, e.g., Ref.~\cite{Mathur:2005zp} for the review)
and firewalls~\cite{Braunstein:2009my,Almheiri:2012rt}  
change the structure of BH spacetime even outside the horizon. 
Also, an interesting class of singularity and ghost free theories
of gravity has been proposed by Ref.~\cite{Modesto:2011kw,Biswas:2011ar}.

In this paper, we consider binary black hole (BBH) systems,
and use gravitational wave (GW) observations as a tool 
to test whether the newly formed black hole genuinely 
behaves like the one predicted by GR or not.
There are various methods proposed for testing GR by means of 
quasinormal mode (QNM) GWs (see an extensive review~\cite{Berti:2015itd}),
for example, tests of the no-hair theorem combining two or more
modes~\cite{Dreyer:2003bv}. QNMs dominate the GWs at 
the ringdown phase of BBH mergers (see also Ref.~\cite{Berti:2009kk}).
In Ref.~\cite{Hughes:2004vw},
testing Hawking's area theorem~\cite{Hawking:1971tu}
has been discussed, which is possible if 
we can determine the masses and spins of BHs
before and after merger independently with a sufficiently high accuracy. 

One of the methods that we propose in this paper 
is the following simple one. 
First, we extract the binary parameters of BBHs by 
taking correlation with the post-Newtonian (PN)
templates~\cite{Blanchet:2013haa, Schafer:2009dq}. 
We assume that we know sufficiently high PN-order 
terms to describe the inspiral phase well. 
Thanks to the development in numerical relativity 
(NR)~\cite{Pretorius:2005gq, Campanelli:2005dd, Baker:2005vv}, 
now we can use simulation results to 
describe the BBH merger phase, deriving accurate gravitational waveforms.
Next, if GR is correct, after the merger phase
we will observe ringdown (QNM) GWs from the remnant BHs
(see, e.g., Ref.~\cite{Berti:2009kk} for a review of the QNMs).
If we do not detect the QNMs as expected,
it is possible to distinguish
the remnant object from the BHs that are predicted by GR
within the assumptions mentioned above. 
It should be noted that our approach is similar to
Ref.~\cite{Luna:2006gw}, in which the authors discussed 
the improvement in parameter estimation 
by combining inspiral and ringdown GWs from compact binaries.  
By contrast, the focus of our work is on the test of GR. 

The other method shown in this paper is even simpler.
When we focus on the dominant QNM, there is a forbidden parameter region
in GR. Just using the ringdown GWs, we can directly discuss
whether the QNM from the remnant compact object is 
consistent with the one from a BH predicted by GR or not. 

This paper is organized as follows.
In Sec.~\ref{sec:prep}, we summarize our tools,
the inspiral and ringdown waveforms from BBHs, 
the fitting formulas for the remnant mass and spin,
and the matched filtering and parameter estimation in the GW data analysis.
In Sec.~\ref{sec:STGR}, two simple tests of GR are presented.
One is to use only the ringdown GWs, and the other is the combination
of inspiral and ringdown phases.
Finally, we summarize and discuss our approach in Sec.~\ref{sec:SD}.
In this paper, we use the geometric unit system, where $G=c=1$,
and the characteristic scale is
$1 \msun = 1.477 {\rm km} = 4.926 \times 10^{-6} {\rm s}$.

%%%%%%%%%%%%%%%%%%%%%%%%%%%%%%%%%%%%%%%%
\section{Preparation}\label{sec:prep}
%%%%%%%%%%%%%%%%%%%%%%%%%%%%%%%%%%%%%%%%

%%%%%%%%%%
\subsection{Target of gravitational waves}
%%%%%%%%%%

According to Kinugawa {\it et al}.~\cite{Kinugawa:2014zha,Kinugawa:2015},
typical total and chirp masses
for Pop III BBHs are $\sim 60 \msun$ and $\sim 30 \msun$, respectively.
Here, the chirp mass of a binary is defined by ${\cal M}= M \eta^{3/5}$
with the total mass $M=m_1+m_2$
and the symmetric mass ratio $\eta=m_1 m_2/M^2$.
This means that $\eta \sim 1/4$ 
for almost equal mass BBHs, which we think typical ones.
In the following discussion, we focus on equal mass BBHs.
Although spins of BBHs can be important, we ignore them here 
for the following reason. 
If we take into account the spins, one may think that the 
accuracy of parameter estimation might be significantly reduced 
due to the degeneracy among the orbital parameters. 
However, in that case the orbital precession induced by the spin 
effects modulates the gravitational waveform. Therefore, 
to a certain extent, this additional information can 
compensate the loss of accuracy due to the degeneracy. 
Hence, for simplicity, we use only the nonspinning
inspiral waveform.

The inspiral phase of GWs from BBHs 
has been extensively studied using the 
PN approximation~\cite{Blanchet:2013haa}.
If we adopt the stationary phase approximation (SPA)~\cite{Damour:2000zb},
we can easily transform the waveform 
into the expression in the frequency domain 
as $\tilde A_{\ell m} e^{i \psi_{\ell m}}$. 
Here, we discuss only the $(\ell=2,\,m=2)$ mode,
and the phase is written as
\bea
 \psi_{22} (v) &= 2 \,\displaystyle{\frac{t_c}{M}} v^3 - 2\Phi_c 
- \displaystyle{\frac{\pi}{4}} 
+ \displaystyle{\frac{3}{128\, \eta \, v^5}} \left[ 1 + {\cal O}(v^2) \right] \,,
\eea
where $v = (M \pi f)^{1/3}$, $t_c$ and $\Phi_c$
are the time and the phase of coalescence,
and the higher-order PN terms are summarized, e.g.,
in Eq.~(A.21) of Ref.~\cite{Brown:2007jx}.
The appropriate SPA amplitude in the frequency domain is deduced
from the time domain description $A_{22}$ by
\bea
\tilde A_{22} = A_{22} \sqrt{\frac{\pi M}{3 v^2 \dot v}} \,,
\eea
where $\dot v$ is given in Eq.~(A.15) of Ref.~\cite{Brown:2007jx}.

After passing the innermost stable circular orbit (ISCO),
the BBHs swiftly plunge to merge. 
Therefore, we terminate the inspiral GW analysis
at the GW frequency for the $(m=2)$ mode at 
ISCO, $f_{\rm ISCO}=(6^{3/2}\pi M)^{-1}$~\cite{Cutler:1994ys}. 
For a typical case with $M=60 \msun$, $\eta=1/4$,
this ISCO frequency is given by $f_{\rm ISCO} = 73.28$Hz.

We can discuss the waveform from 
the merger phase accurately using NR
simulations~\cite{Pretorius:2005gq, Campanelli:2005dd, Baker:2005vv}.
The whole of GW waveforms from BBH coalescence
are also well modeled in the effective-one-body approach
(see, e.g., Ref.~\cite{Taracchini:2013rva} for the latest development).
However, here, we do not make use of the GWs from the merger phase. 
There is much progress in the understanding of
the mass, spin and recoil velocity of the remnants after BBH mergers
which allows us to connect the observation of the inspiral phase
to the ringdown phase 
(see, e.g., Ref.~\cite{Healy:2014yta} for the latest formulas).
Here, we use the formulas for initially nonspinning cases.
The phenomenological fitting formulas for the remnant mass
and spin are given by~\cite{Healy:2014yta}
\bea
\frac{M_{\rm rem}}{M} &=& (4\eta)^2\,
\left( M_0 + K_{2d}\,\delta m^2 + K_{4f}\,\delta m^4 \right)
+ \left[1 + \eta (\tilde{E}_{\rm ISCO}+11) \right] \delta m^6 \,,
\label{eq:Mrem_formula}
\\
\alpha_{\rm rem} &=& \frac{S_{\rm rem}}{M^2_{\rm rem}}
= (4\eta)^2 \left( L_0 + L_{2d}\,\delta m^2 + L_{4f}\,\delta m^4 \right)
+ \eta \tilde{J}_{\rm ISCO} \delta m^6 \,,
\label{eq:arem_formula}
\eea
where 
$\delta m = (m_1-m_2)/M$ ($=-\sqrt{1-4\eta}$ for $m_1<m_2$)
and $\tilde{E}_{\rm ISCO}$ and $\tilde{J}_{\rm ISCO}$ are
the specific energy and angular momentum at ISCO in the test particle 
approximation (see, e.g., Ref.~\cite{Ori:2000zn}).
$M_0$, $K_{2d}$, $K_{4f}$, $L_0$, $L_{2d}$ and $L_{4f}$
are the fitting parameters summarized in Table VI of Ref.~\cite{Healy:2014yta}.
$\alpha$ is $a/M$ of the Kerr BH with the mass $M$ and Kerr parameter $a$.
More specifically, for equal mass cases, i.e., 
$\eta=1/4$ and $\delta m=0$, we have
\bea
\frac{M_{\rm rem}}{M} &=& 0.951507 \pm 0.000030 \,,
\cr
\alpha_{\rm rem} &=& 0.686710 \pm 0.000039 \,,
\label{eq:fittingS}
\eea
including the magnitude of numerical errors.
As we noted before, 
the remnant mass becomes $M_{\rm rem} = 57.0904 \msun$ 
for a representative case with $M=60 \msun$, $\eta=1/4$. 

The above formulas obtained by fitting the results of BBH simulations
in the case of nonprecessing BBHs 
have $1\%$ relative error, which is mainly caused by the 
extraction of the GW radiation at a finite radius 
and finite mesh resolution in the NR simulations.
The radial extrapolation errors will be reduced by using
a perturbative extraction method~\cite{Nakano:2015rda,Nakano:2015pta}.
Also for precessing BBHs, we may have much larger errors.
Although these errors are directly related to the following analysis,
we expect that the fitting formulas will be improved by more NR simulations.
Therefore, we just ignore them in the following analysis. 

Using the estimated remnant BH's mass and spin, 
we discuss the ringdown phase.
The waveform is modeled as
\bea
h(f_c,\,Q,\,t_0,\,\phi_0;\,t) =
\begin{cases}
e^{ - \frac {\pi \,f_c\,(t-t_0)}{Q}}\,\cos(2\,\pi \,f_c\,(t-t_0)-\phi_0) 
& {\rm for} \quad t \geq t_0 \,, \\
0 & {\rm for} \quad t < t_0 \,,
\end{cases}
\eea
where $t_0$ and $\phi_0$ are the initial ringdown time and phase, respectively. 
The central frequency $f_c$ and the quality factor $Q$
are related to the real ($f_R$) and imaginary ($f_I$) parts
of the QNM frequency as
\bea
f_R = f_c \,, \quad f_I = - \frac{f_c}{2Q} \,,
\label{eq:freqRI}
\eea
which depend on the harmonics index $(\ell,\,m)$ and the overtone index $n$.
Here, we focus on the dominant $(\ell=m=2)$ least-damped $(n=0)$ mode
and the fitting formulas for $f_c$ and $Q$ are given in Ref.~\cite{Berti:2005ys} as
\bea
f_c &=& \frac{1}{2\pi M_{\rm rem}}
\left[ 1.5251 - 1.1568 (1-\alpha_{\rm rem})^{0.1292} \right] 
\cr 
&=& 538.4 \left( \frac{M}{60 \msun} \right)^{-1}
\left[ 1.5251 - 1.1568 (1-\alpha_{\rm rem})^{0.1292} \right]
\, [{\rm Hz}] \,,
\label{eq:Ffitting}
\\
Q &=& 0.7000 + 1.4187 (1-\alpha_{\rm rem})^{-0.4990} \,.
\label{eq:Qfitting}
\eea
For the fiducial values, $M=60 \msun$, $\eta=1/4$, 
we have 
$M_{\rm rem}=57.0904 \msun$ and $\alpha_{\rm rem}=0.686710$, 
and the above formulas derived based on GR predict
$f_c = 299.5$Hz and $Q=3.232$ for the ringdown GW. 
Here, it is noted that the fitting formulas in Eqs.~\eqref{eq:Ffitting}
and~\eqref{eq:Qfitting} have $2\%$ and $1\%$ errors, respectively.
Therefore, although we use the fitting formulas for simplicity in this paper,
we should use the original data
in Ref.~\cite{Berti_QNM} for the strict analysis.

%%%%%%%%%%
\subsection{Matched filtering and parameter estimation}

To analyze the GWs from the inspiral and ringdown phases,
we use the matched filtering method because the waveforms are known well.
Using the inner product,
\bea
\langle a|b \rangle = 4 \Re \int_0^{\infty} \frac{\tilde a (f) \tilde b^* (f)}{S_n(f)} df \,,
\label{eq:inner}
\eea
where $S_n(f)$ denotes
the power spectral density of GW detector's noise,
the optimal signal-to-noise ratio (SNR) for a waveform $h$
is given by
\bea
\snr &=& \langle h|h \rangle^{1/2} 
\cr
&=&
2 \left[\int_0^{\infty} \frac{|\tilde h (f)|^2}{S_n(f)} df \right]^{1/2}
\,.
\eea

\begin{figure}[!ht]
\begin{center}
 \includegraphics[width=0.5\textwidth]{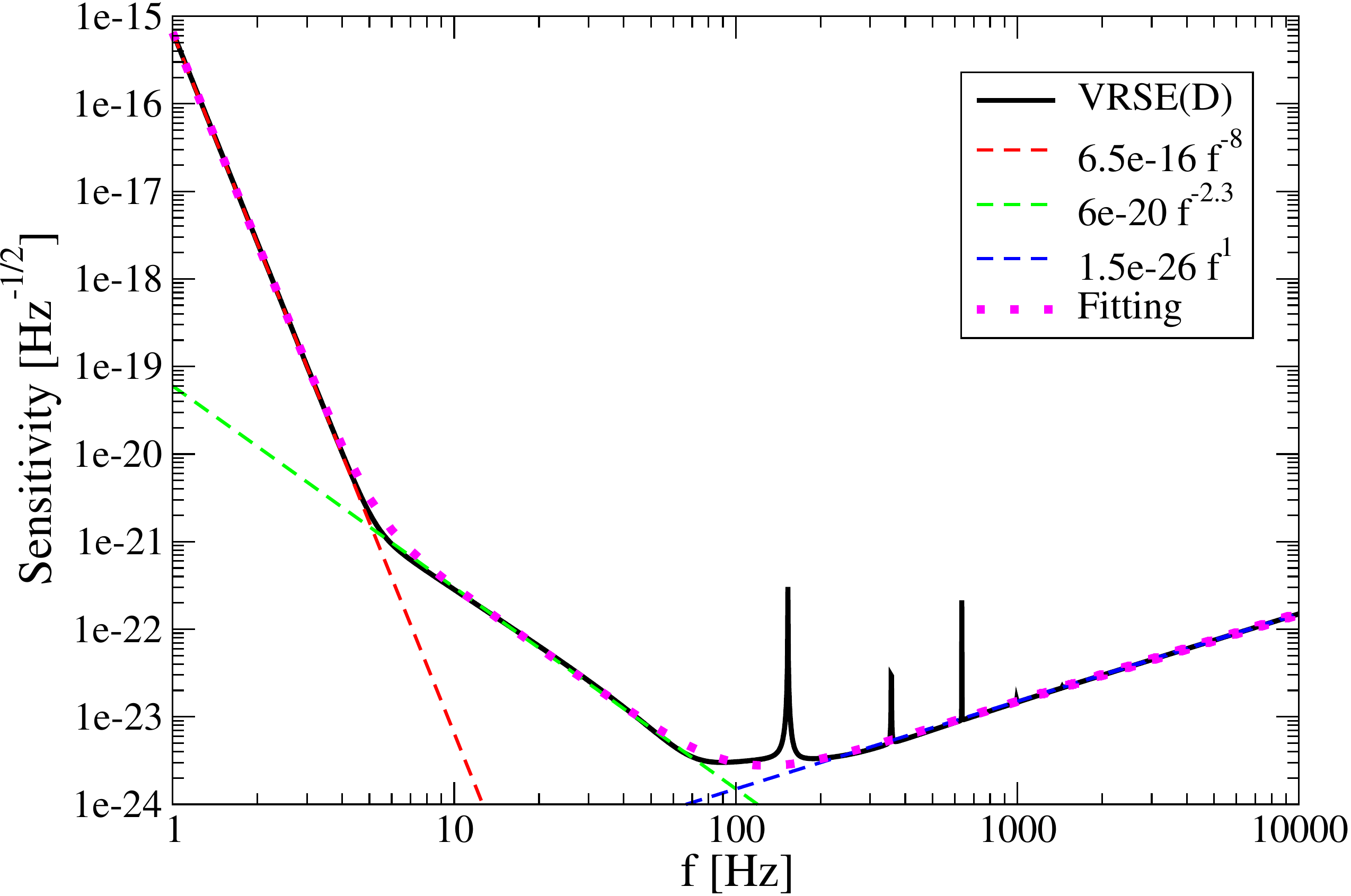}
\end{center}
 \caption{Fitting curve based on the sensitivity curve of KAGRA [bKAGRA, VRSE(D) configuration]
shown in Ref.~\cite{bKAGRA}.}
 \label{fig:bKAGRA}
\end{figure}

We assume a single GW detector, KAGRA~\cite{Somiya:2011np,Aso:2013eba}, here.
In Fig.~\ref{fig:bKAGRA}, we show the expected noise curve of KAGRA
[bKAGRA, VRSE(D) configuration] presented in Ref.~\cite{bKAGRA}, 
which can be fit well by 
\bea
S_n(f)^{1/2} = 10^{-26} \left( 6.5\times 10^{10} f^{-8}
+ 6\times 10^6 f^{-2.3}
+ 1.5 f^1 \right) \, [{\rm Hz^{-1/2}}] \,,
\label{eq:fittingNC}
\eea
where the frequency $f$ is in units of Hz.
Of course, we can discuss the other detectors
(Advanced LIGO~\cite{TheLIGOScientific:2014jea},
Advanced Virgo~\cite{TheVirgo:2014hva}, GEO-HF~\cite{2014CQGra..31v4002A},
and so on) just by changing $S_n(f)$.

To calculate the parameter estimation errors
for the inspiral and ringdown GWs,
we use the Fisher information matrix,
\bea
\Gamma_{ij} = \left<
\frac{\partial h}{\partial \theta^i} \left| 
\frac{\partial h}{\partial \theta^j}
\right> \right|_{\theta=\theta_{\rm true}} \,,
\eea
where $\theta^i$ is the parameters of the waveforms
and $\theta_{\rm true}$ denotes the true values
of the parameters of the source.
Then, the rms errors
in the estimated parameters and the covariance between
two parameters are derived by the inverse matrix
$(\Gamma^{-1})^{ij}$ as
\bea
(\Delta \theta^i)_{\rm RMS} &=& \sqrt{(\Gamma^{-1})^{ii}} \,,
\cr
c_{ij} &=& 
\frac{(\Gamma^{-1})^{ij}}{\sqrt{(\Gamma^{-1})^{ii}(\Gamma^{-1})^{jj}}}
\,.
\eea
Here, we do not sum over $i$ and $j$.
$(\Delta \theta^i)_{\rm RMS}$ scales as $1/\snr$.

For the inspiral phase, 
we calculate the parameter estimation errors for
$\{M,\,\eta,\,t_c,\,\Phi_c\}$, 
Here, we use the total mass instead of the chirp mass 
for the parametrization of the inspiral signal, 
simply because the fitting formulas
for the remnant mass and spin are written in terms of $M$ and $\eta$.
To evaluate the inner product~\eqref{eq:inner}, 
we take the integration range 
between $10$Hz and $f_{\rm ISCO}$, 
For the ringdown phase, we discuss the parameter estimation 
with respect to  
$\{f_c,\,Q,\,t_0,\,\phi_0\}$,
and the frequency interval for the integration
is between $10$ and $2500$Hz.

We should note that in practice
the location of the GW source in the sky
and the GW polarization angle in a detector frame are also
the parameters to describe the GW signals.
For example, Ajith and Bose~\cite{Ajith:2009fz} 
estimated the parameter errors of BBHs
in a single detector or a detector network
for the case of the complete set of parameters.
This direction to discuss more precise parameter estimation
is one of our future studies.

%%%%%%%%%%%%%%%%%%%%%%%%%%%%%%%%%%%%%%%%
\section{Simple test of GR}\label{sec:STGR}
%%%%%%%%%%%%%%%%%%%%%%%%%%%%%%%%%%%%%%%%

According to Ref.~\cite{LCGT:2011aa},
individual SNRs for the inspiral and ringdown phase signals
are comparable for a gravitational wave detector, KAGRA.
when the total BBH mass ($\sim$ remnant BH mass) is $\sim 60 \msun$. 
Since there is a difficulty in determining the initial ringdown amplitude
due to the ambiguity of the initial time, for simplicity,
we set the SNRs for the inspiral and ringdown phases
to be equal for the typical case (with $M=60 \msun$ and $\eta=1/4$
for inspiral and $M_{\rm rem}=57.0904 \msun$ and
$\alpha_{\rm rem}=0.686710$ for ringdown).
The assumption of the same SNR for the inspiral and ringdown phases
is just for simplicity, 
and we can apply the following analysis for general SNR cases.
The information of SNRs is imprinted in the Fisher information matrix of each phase.
We briefly discuss the effect by setting different SNRs 
for the inspiral and ringdown phases in Sec.~\ref{sec:SD}.

%%%%%%%%%%
\subsection{Only ringdown}\label{sec:OR}
%%%%%%%%%%

\begin{figure}[!ht]
\begin{center}
 \includegraphics[width=0.5\textwidth]{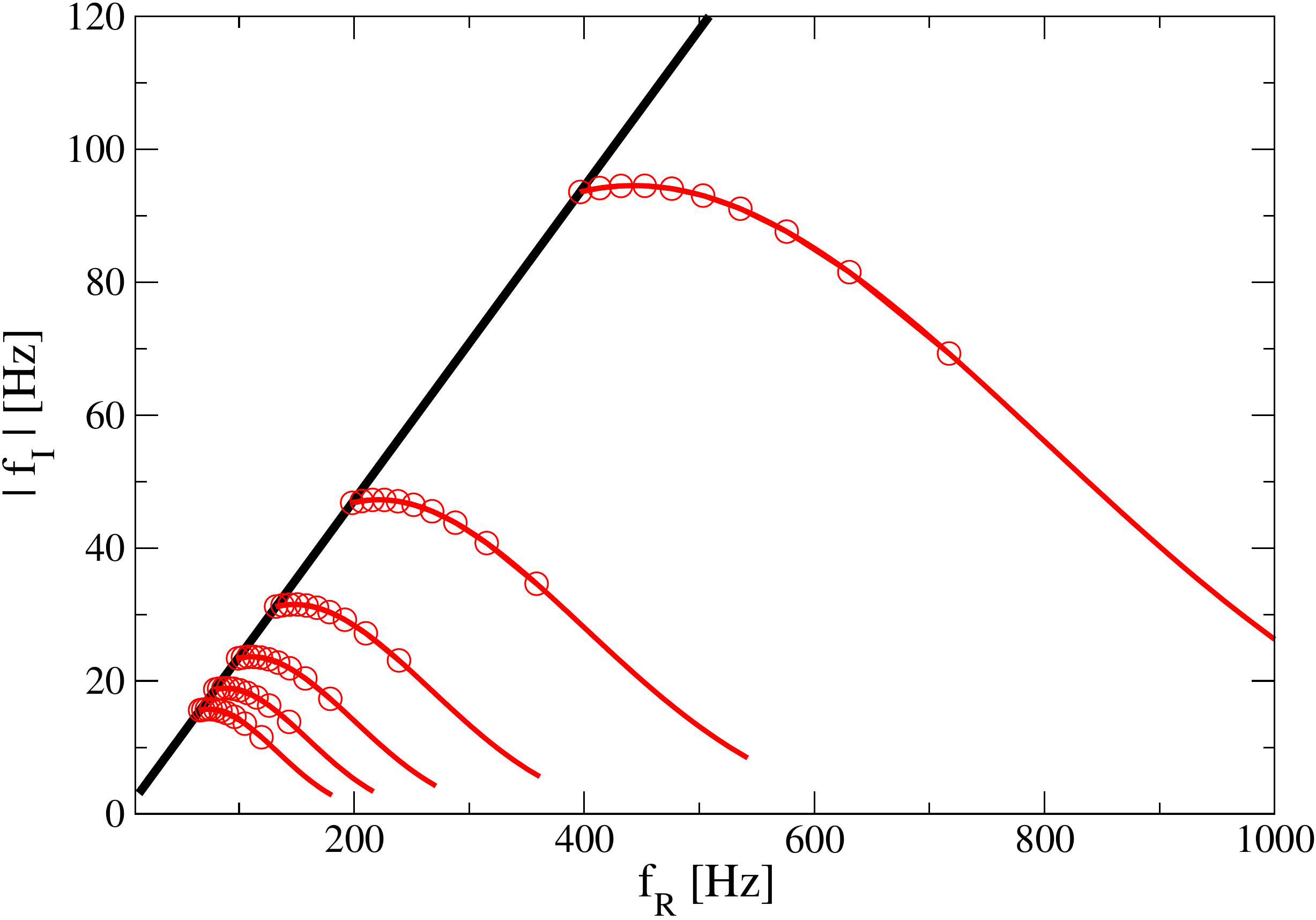}
\end{center}
 \caption{Real ($f_R$) and imaginary ($f_I$) parts of QNM frequencies
for the dominant $(\ell=2,\,m=2)$ least-damped $(n=0)$ mode.
The (black) thick line shows the Schwarzschild limit,
and the (red) curves are for various mass cases terminated
at the spin $\alpha=0.998$~\cite{Thorne:1974ve}.
From the top of the (red) curves, we are considering BH masses,
$M/\msun=30,\,60,\,90,\,120,\,150$, and $180$, respectively.
The (red) circles for each line denote the spin dependence
$\alpha=0,\,0.1,\,0.2,\,0.3,\,0.4,\,0.5,\,0.6,\,0.7,\,0.8$,
and $0.9$ from the left.}
 \label{fig:prohibit}
\end{figure}

First, using only the ringdown GWs, we propose a simple method to test 
whether the compact object emitting the ringdown GWs
is a BH predicted by GR or not.
Figure~\ref{fig:prohibit} shows
the QNM frequencies for the dominant $(\ell=2,\,m=2)$ 
least-damped $(n=0)$ mode in the $(f_R,\,f_I)$ plane. 
In GR, the top-left side of the thick black line 
is prohibited. The boundary thick black line 
corresponds to the Schwarzschild limit, 
which is obtained 
by setting $\alpha_{\rm rem}=0$, i.e., 
\bea
\frac{|f_I|}{f_R} \approx 0.236 \,, 
\eea
in Eqs.~\eqref{eq:freqRI}, 
\eqref{eq:Ffitting} and \eqref{eq:Qfitting}. 
In principle, if we obtain the parameters
in the forbidden region from GW observations,
we can conclude that the compact object is not the one 
predicted by GR.

\begin{figure}[!ht]
\begin{center}
 \includegraphics[width=0.48\textwidth]{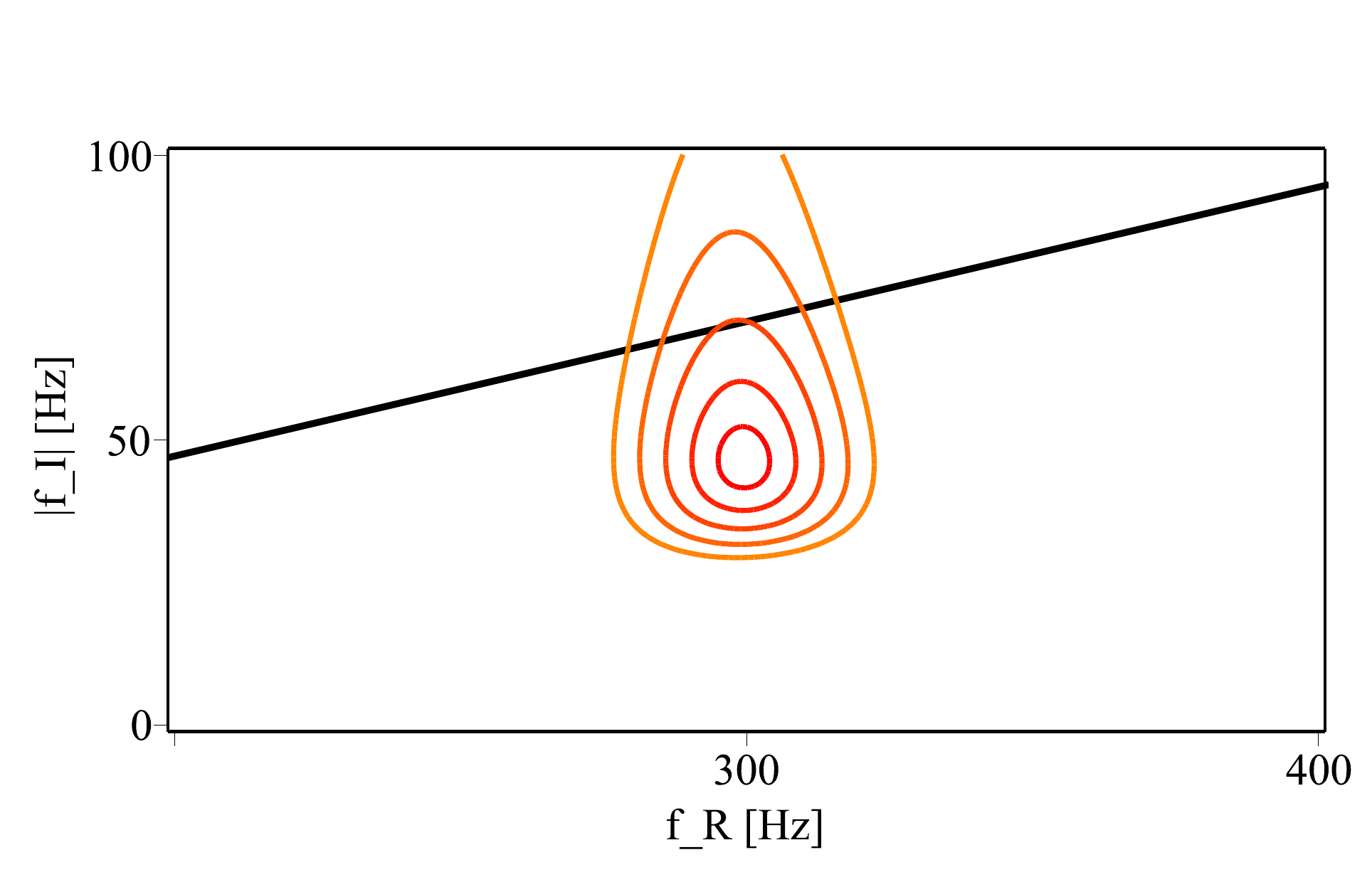}
 \includegraphics[width=0.48\textwidth]{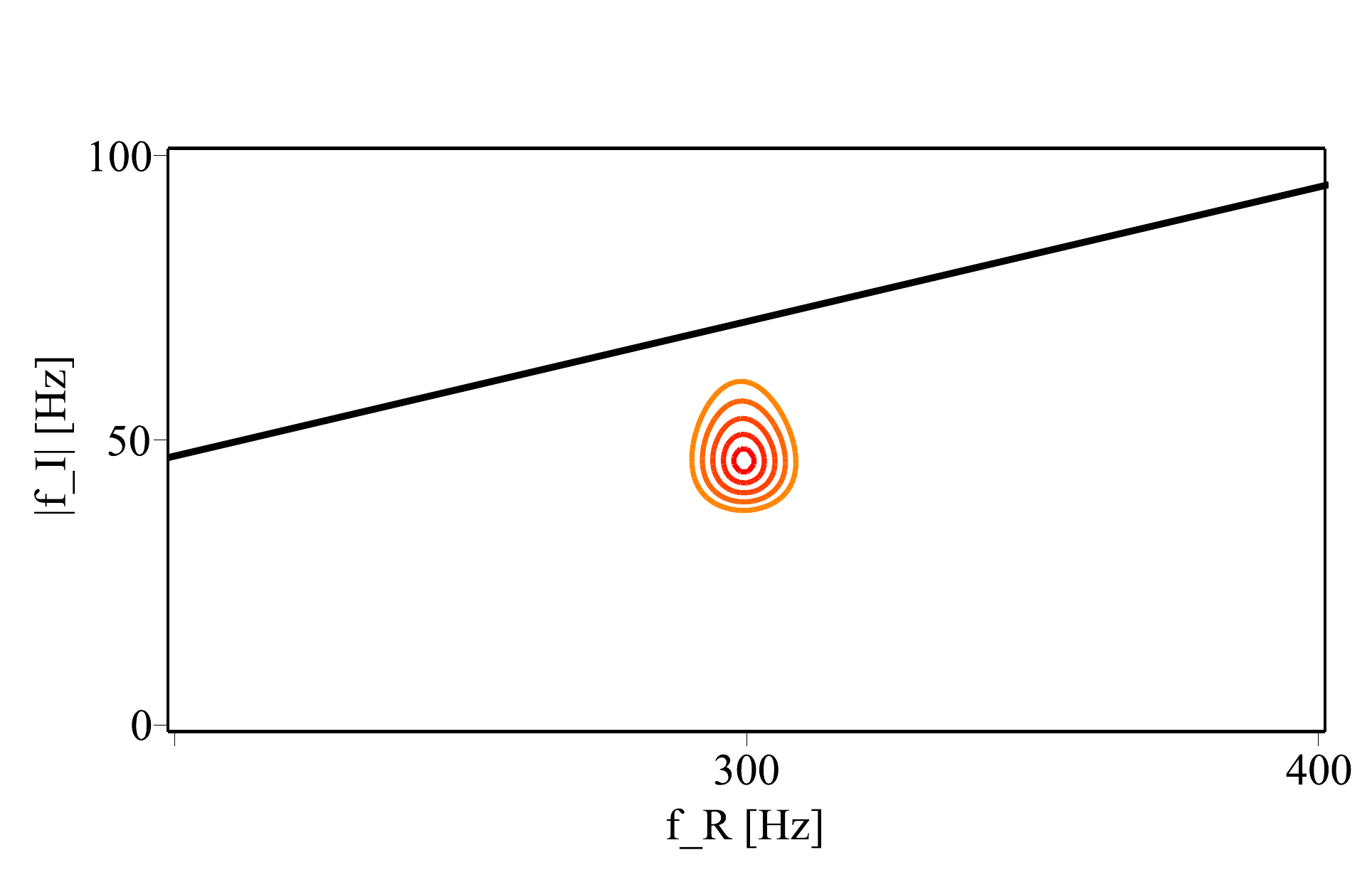}
\end{center}
 \caption{In the ($f_R,\,f_I$) plane, the left and right panels show 
the parameter estimation in the cases with $\snr=20$ and $50$
for the typical case
(with $M_{\rm rem}=57.0904 \msun$ and $\alpha_{\rm rem}=0.686710$), respectively.
The (black) thick line shows the Schwarzschild limit which is same
as that in Fig.~\ref{fig:prohibit},
and the ellipses are the contours of
$1\sigma,\,2\sigma,\,3\sigma,\,4\sigma,$ and $5\sigma$.
Here, the time and phase parameters ($t_0,\,\phi_0$) have been marginalized out.}
 \label{fig:sample}
\end{figure}

However, in practice, 
there are parameter estimation errors in the GW data analysis.
For our typical example with $f_c = 299.5{\rm Hz},\,Q=3.232,\,t_0=0$,
and $\phi_0=0$
[$f_R = 299.5{\rm Hz}$ and $f_I=-46.34{\rm Hz}$ from Eqs.~\eqref{eq:freqRI}],
we show the contours of the parameter estimation errors 
in Fig.~\ref{fig:sample}.
Here, since we do not discuss the errors of $t_0$ and $\phi_0$,
we integrated the probability distribution over both 
$t_0$ and $\phi_0$~\cite{Finn:1992wt}.
In our typical case, the expected errors are sufficiently 
small to fit the ringdown GW with $\snr=50$ within the QNM parameter region
allowed in GR at the $5\sigma$ level.
On the other hand, the error circle for the signal with $\snr=20$ is not sufficiently
small in this sense at that level, while it is small enough for 3$\sigma$ level arguments. 
Here, $5\sigma$ ($3\sigma$) denotes that for the bidimensional
(Rayleigh) distribution, which means that the probability falling in the 
$5\sigma$ ($3\sigma$) circle is about $1-3.7\times 10^{-6}$ ($1-1.1\times 10^{-2}$)
since the distribution has 2 degrees of freedom.
[In the case of the ordinary one-dimensional Gaussian distribution, 
the probability falling in the $5\sigma$ ($3\sigma$) region is about
$1-5.7\times 10^{-7}$ ($1-2.7\times 10^{-3}$).]

\begin{figure}[!ht]
\begin{center}
 \includegraphics[width=0.48\textwidth]{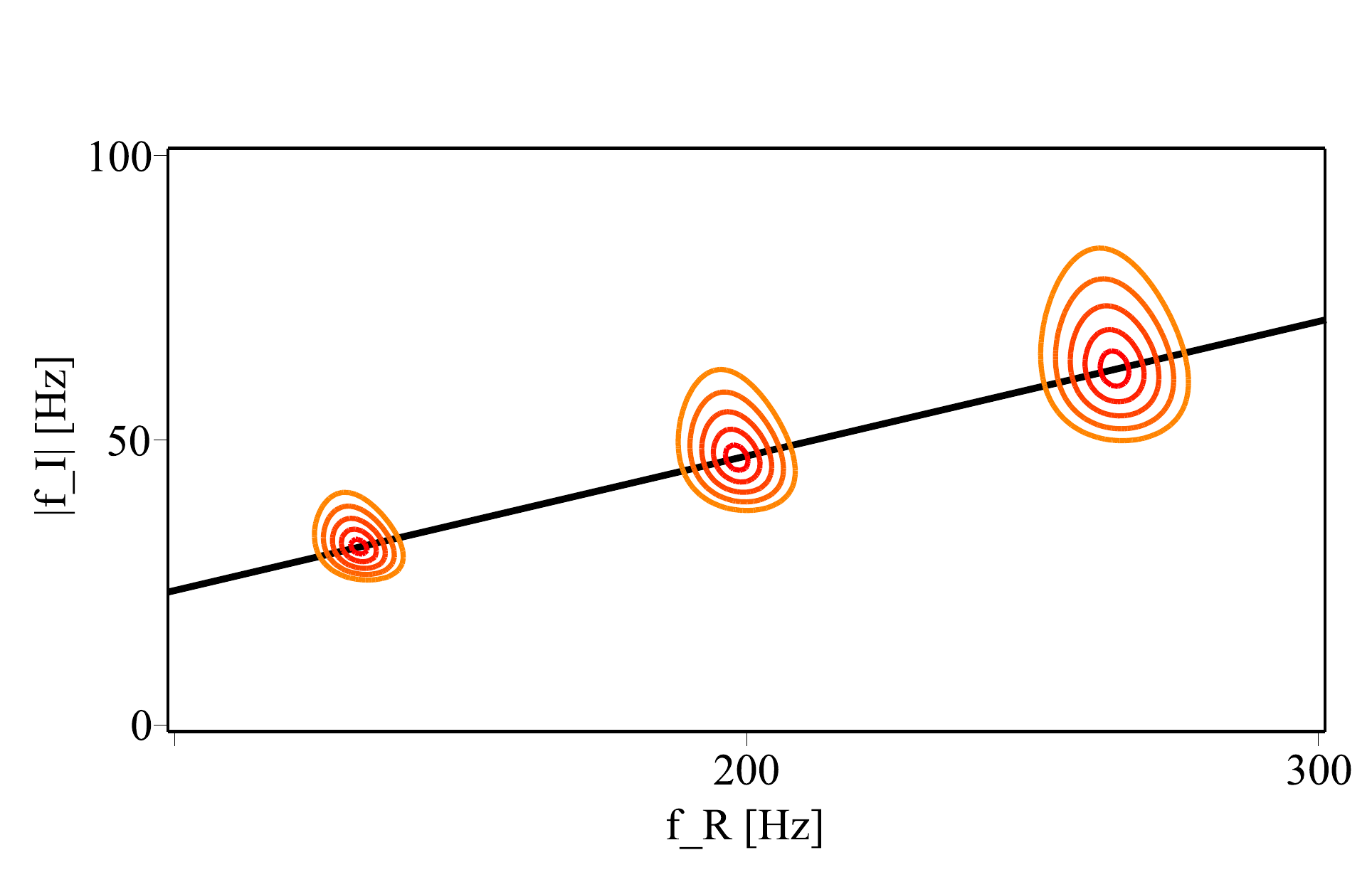}
\end{center}
 \caption{In the ($f_R,\,f_I$) plane, this figure shows 
the parameter estimation in the cases with $\snr = 50$
for a Schwarzschild black hole with
$M_{\rm rem}/\msun=45$ (right), $60$ (center) and $90$ (left).
The (black) thick line shows the Schwarzschild limit which is same
as that in Fig.~\ref{fig:prohibit},
and the ellipses are the contours of
$1\sigma,\,2\sigma,\,3\sigma,\,4\sigma,$ and $5\sigma$.
Here, the time and phase parameters ($t_0,\,\phi_0$) have been marginalized out.}
 \label{fig:sample_Sch}
\end{figure}

To discuss the region prohibited by GR, 
we present the parameter estimation for the Schwarzschild
($\alpha_{\rm rem}=0$) case in Fig.~\ref{fig:sample_Sch}.
Here, we fixed $\snr = 50$ and considered the remnant masses,
$M_{\rm rem}/\msun = 45$, $60$, and $90$. From the $5\sigma$ contours,
there is an upper bound of the GR prediction for $|f_I|/f_R$, 
and we find that the region of $|f_I|/f_R > F_{\rm max}$
for each mass case is rejected by GR.
Here, $F_{\rm max}$, which denotes the maximum of $|f_I|/f_R$
allowed in GR,
is $0.321$ (for $M_{\rm rem}=45 \msun$),
$0.320$ ($60 \msun$) and 
$0.316$ ($90 \msun$) for $\snr = 50$.
If NR simulations for the extreme spinning BBH are available,
we can also give the lower bound of the GR prediction for $|f_I|/f_R$.

It is noted that a powerful method to find
ringdown signals in multiple GW detectors has been proposed 
by Talukder, Bose, Caudill and Baker~\cite{Talukder:2013ioa}.
Although we have considered the above GW data analysis with a single detector,
we may expect a better parameter estimation in a detector network.

%%%%%%%%%%
\subsection{Consistency analysis with inspiral and ringdown}
%%%%%%%%%%

Next, we propose a consistency test by combining the data 
from inspiral and ringdown GWs.
We use the PN waveform for the inspiral phase to extract 
the binary parameters, 
and the formulas in Eqs.~\eqref{eq:Mrem_formula} and \eqref{eq:arem_formula}
of Sec.~\ref{sec:prep} are applied to obtain 
the GR prediction for the parameters of the remnant black hole.
Then, we can present the QNM frequency expected in GR 
in the ($f_R,\,f_I$) parameter space.

To take into account the observational errors in the estimate of 
the expected QNM, 
we assume that the true signal is given by the GR template 
with $\vect{\theta}$,
and the parameters estimated from the 
inspiral and ringdown signals 
are $\vect{\theta}_{\rm Insp}$ and $\vect{\theta}_{\rm Ring}$, respectively.
Here, $\vect{\theta}$ consists of the parameters 
$\{f_c,\,Q,\,t_0,\,\phi_0\}$, 
which are commonly used for the ringdown GW data analysis.
For the ringdown phase, we treat the above parameters
to calculate the parameter estimation errors and
assume the Gaussian distribution for the parameters.
In the inspiral-phase analysis, we use another set of parameters
$\tilde{\vect{\theta}} = \{M,\,\eta,\,t_c,\,\Phi_c\}$.

Here, it is useful to have the relation between the inspiral parameters $\tilde{\vect{\theta}}$
and the ringdown parameters $\vect{\theta}$ as fitting functions. From Eq.~\eqref{eq:arem_formula},
we have
\bea
\alpha_{\rm rem} &=& 0.830028 \left(\frac{\eta}{0.25}\right) 
- 0.143761 \left(\frac{\eta}{0.25}\right)^2
+ 0.00180831 \left(\frac{\eta}{0.25}\right)^{12} \,. 
\eea
The above relation gives one-to-one mapping 
in the parameter ranges, 
$0 \leq \eta \leq 0.35282872$ and $0 \leq \alpha_{\rm rem} \leq
0.99800367$. 
It is noted that, although $\eta >0.25$ is an unphysical value,
we allow the values here. 
Combining the above equation with Eq.~\eqref{eq:Qfitting},
we find that $\eta$ is fitted as a function of $Q$ to obtain 
\bea
\eta &=& 0.353039 - \frac{0.208266}{Q^2} - \frac{10.9583}{Q^4} - \frac{21.4540}{Q^6} \,.
\eea 
The restriction on the parameter space 
to keep the one-to-one mapping 
becomes $2.11870 \leq Q \leq 32.2555$.
The decay time is calculated as $Q/(\pi f_c)$.
Using Eqs.~\eqref{eq:Mrem_formula} and \eqref{eq:Ffitting}
(and also the above fitting functions for $\alpha_{\rm rem}$ and $\eta$),
the total mass $M$ in the inspiral phase is written by $f_c$ and $Q$ as
\begin{eqnarray}
M  &=& \frac{1}{f_c}
\left[
- 0.0434932 - 0.127430 \ln \left( \frac{1}{Q} + 0.163772 \right)
+ \frac{0.0646167}{\sqrt{Q}} \right] \,.
\eea

To find the expected parameter region of the QNM, 
we use the following simple estimator
(more detailed studies, e.g., by using Markov chain
Monte Carlo methods, will be presented in future):
\bea
F(\vect{\theta})
= {\cal N} \exp 
\left[ - \frac{1}{2}\tilde\Gamma_{ij}^{\rm Insp}
(\tilde\theta^i(\vect{\theta})-\tilde\theta^i(\vect{\theta}_{\rm Insp}))
(\tilde\theta^j(\vect{\theta})-\tilde\theta^j(\vect{\theta}_{\rm Insp}))
- \frac{1}{2}\Gamma_{ij}^{\rm Ring} 
(\theta^i-\theta_{\rm Ring}^i) (\theta^j-\theta_{\rm Ring}^j) \right] \,,
\label{eq:combined}
\eea
where ${\cal N}$ is a normalization constant which we do not take care of,
and $\tilde\Gamma_{ij}^{\rm Insp}$ and $\Gamma_{ij}^{\rm Ring}$ denote
the respective Fisher information matrices after integrating the probability 
distribution over $(t_c,\,\Phi_c)$ and $(t_0,\,\phi_0)$.

The strategy to estimate the parameter region by using
Eq.~\eqref{eq:combined} is as follows:
\begin{itemize}
\item[(1)]
For given $\tilde{\vect{\theta}}(\vect{\theta}_{\rm Insp}) (=:\tilde{\vect{\theta}}_{\rm Insp})$
(in practice, we give $\tilde{\vect{\theta}}_{\rm Insp} = \{M=60 \msun,\,\eta=1/4\}$
and derive $\vect{\theta}_{\rm Insp}$),
we calculate $\tilde\Gamma_{ij}^{\rm Insp}$ with the bKAGRA noise curve.

\item[(2)]
Assuming the narrow ringdown signal in the frequency domain,
we prepare $\Gamma_{ij}^{\rm Ring}$ for the white noise (analytically).

\item[(3)]
For given $\vect{\theta}_{\rm Ring}$ (and $\Gamma_{ij}^{\rm Ring}$ for it),
we find the maximum of Eq.~\eqref{eq:combined} by
\bea
\frac{\partial F(\vect{\theta})}{\partial \theta^i} = 0 \,.
\eea

\item[(4)]
Inserting the solution of the above equation
$\vect{\theta} = \{f_c,\,Q\}$ back into Eq.~\eqref{eq:combined},
we check whether the situation with the parameters 
$(\vect{\theta}_{\rm Insp},\,\vect{\theta}_{\rm Ring},\,\vect{\theta})$ is
in the $5\sigma$ level of the detector noise realization or not.
\end{itemize}

Here, $5\sigma$ denotes that the value in the exponent in 
Eq.~\eqref{eq:combined} becomes $-5^2/2$, which means that 
the probability falling in the 
$5\sigma$ circle is about $1-5\times 10^{-5}$ 
since the distribution has 4 degrees of freedom.  
Employing our fiducial values, $M=60 \msun$, $\eta=1/4$,
the expected region of the QNM frequency in the $5\sigma$ level
is shown in Fig.~\ref{fig:combF_60}.
Here, we have fixed $\snr=50$ for both the inspiral and ringdown GWs.
Compared with the right figure of Fig.~\ref{fig:sample},
the allowed region in this figure has a larger extension in the horizontal direction. 
This is due to the parameter estimation errors of the inspiral phase.
We repeat the meaning of this plot. 
Under the condition that we measure the values of both 
$\tilde{\vect{\theta}}_{\rm Insp}$ and 
$\vect{\theta}_{\rm Ring}$, 
we choose the most probable values for the parameters 
$\vect{\theta}$. Assuming that the true values are the 
most probable values as used in the usual Fisher-matrix analysis, 
we evaluate the probability that the detector noise produces 
such a deviation in the measurement of 
$\tilde{\vect{\theta}}_{\rm Insp}$ and 
$\vect{\theta}_{\rm Ring}$. 
The probability that the noise realization falls outside 
the contour is $5\times 10^{-5}$. Therefore, if we find that 
the parameter estimate from the ringdown signal
deviates from the prediction from the inspiral signal exceeding 
the contour in Fig.~\ref{fig:sample}, we can conclude that 
there is something wrong with the GR prediction. 
Here, under an assumption that 
the nonlinearity of GR is correct for the inspiral and merger phases,
it is possible to distinguish
the remnant object from the BHs that are predicted by GR.

\begin{figure}[!ht]
\begin{center}
 \includegraphics[width=0.5\textwidth]{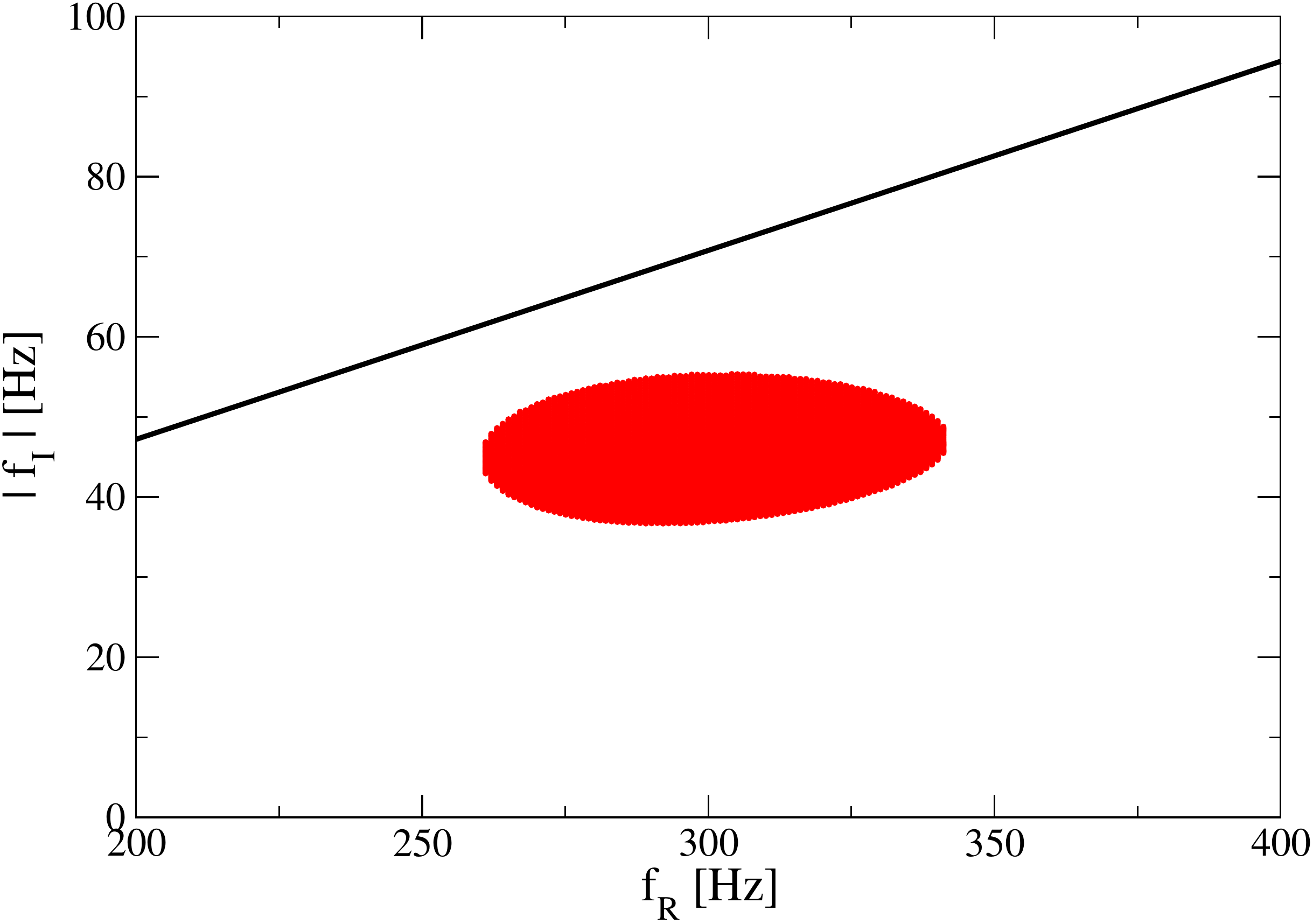}
\end{center}
 \caption{The QNM frequency expected in GR ($5\sigma$ level) from
the inspiral phase with the total mass $M=60 \msun$ 
and symmetric mass ratio $\eta=1/4$ [the (red) filled region].
The (black) thick line shows the Schwarzschild limit which is same as
that in Fig.~\ref{fig:prohibit}.}
\label{fig:combF_60}
\end{figure}

%%%%%%%%%%%%%%%%%%%%%%%%%%%%%%%%%%%%%%%%
\section{Summary and Discussion}\label{sec:SD}
%%%%%%%%%%%%%%%%%%%%%%%%%%%%%%%%%%%%%%%%

In this paper, we mainly focused on 
a specific BBH with the total mass $M=60 \msun$
and the symmetric mass ratio $\eta=1/4$, which 
would be the typical one for Pop III BBHs~\cite{Kinugawa:2014zha,Kinugawa:2015}.
It is found that we can perform meaningful tests of GR, 
assuming that the GW signal has $\snr=50$.
An easy extension of the present study is to treat various total mass cases.
For total masses lower than $M=60 \msun$, we have fewer SNRs
for the ringdown phase than those for the inspiral phase
and expect that a larger elongation in the vertical direction 
in the $(f_R,\,f_I)$ plane because of Fig.~\ref{fig:sample}.
On the other hand, for total masses higher than $M=60 \msun$,
we will have a larger elongation in the horizontal direction.
We also need to discuss various mass ratios and spins in the inspiral phase.
The statistical treatment will be also improved in our future work.

In Fig.~\ref{fig:combF_60}, 
we have observed that the expected region shows a 
large elongation in the horizontal direction. 
This is due to the parameter estimation errors for the inspiral signal
and, more specifically, originates from marginalizing $t_c$ and $\Phi_c$
in the probability distribution.
The parameter estimation errors of $t_c$ and $\Phi_c$ arise from
the short frequency integration interval
between $10$Hz and $f_{\rm ISCO}=73.28$Hz .
The number of GW cycles
during this frequency range is $N_{\rm GW} \approx 30$.
When we change the lower integration bound to $20$Hz,
the situation becomes much worse, i.e.,
the number of GW cycles is just $N_{\rm GW} \approx 6$.

Here, if we can also detect the inspiral phase by using
a space-based GW detector, such as DECIGO~\cite{Seto:2001qf}, 
the situation will improve a lot 
(see, e.g., Ref.~\cite{Nair:2015bga} for the synergy
in the parameter estimation of binary inspirals).
For example, $N_{\rm GW} \approx 5400$ from $0.5$Hz in our specific case.
Therefore, even if we assume the same SNR for the inspiral phase,
the parameter estimation of $M$ and $\eta$ 
and the QNM prediction will be very precise.

Kinugawa {\it et al}.~\cite{Kinugawa:2015} showed
that the expected detection  rate of BH-BH mergers by KAGRA
with typical total mass $\sim 60 \msun$ is given by 
\begin{equation}
262 \,{\rm events~yr^{-1} (SFR_P/(10^{-2.5}~\msun \rm~yr^{-1}~Mpc^{-3}))\cdot Err_{sys}} \,,
\end{equation}
where $\rm SFR_{p}$ and $\rm Err_{sys}$ are the peak value 
of the Pop III star formation rate and the systematic error 
with $\rm Err_{sys}=1$ for their fiducial model, respectively.
They have estimated that ${\rm Err_{sys}}$ ranges from 0.056 to 2.3 
due to the unknown parameters such as the common envelope parameter, 
the kick velocity, and the loss fraction
as well as the unknown distribution functions
such as the initial mass function and the initial eccentricity function.
The minimum value corresponds to the worst model 
in which they adopt the most pessimistic values of the parameters
and distribution functions within the ranges that are likely. 
The factor ${\rm (SFR_P/(10^{-2.5}~\msun \rm~yr^{-1}~Mpc^{-3}))}$
also depends on the models and Kinugawa {\it et al}.~\cite{Kinugawa:2015}
argued that it ranges from 0.019 to 16. 
Therefore, the event rate of Pop III BH-BH mergers 
which will be detected by KAGRA ranges from 
0.28 to 9641 ${\rm yr^{-1}}$.
The event rate for Advanced LIGO and Advanced Virgo will be similar.
Since no such event has been found so far, 
the event rate should be smaller than 1000 ${\rm yr^{-1}}$. 
Adopting a simple geometric mean of this allowed range, 
we have a rough estimate of the expected rate of 500 ${\rm yr^{-1}}$. 
Then, the rates of events with $\snr > 20$ and $\snr > 50$ are 
32 and 2 ${\rm yr^{-1}}$, respectively.
Therefore, there is a good chance to confirm (or refute)
the Einstein theory in the strong gravity regime
by observing the expected QNMs.

%%%%%%%%%%%%%%%%%%%%%%%%%%%%%%%%%%%%%%%
\acknowledgments
%%%%%%%%%%%%%%%%%%%%%%%%%%%%%%%%%%%%%%%

This work was supported by 
the Ministry of Education, Culture, Sports, Science and Technology (MEXT)
Grant-in-Aid for Scientific Research on Innovative Areas,
``New Developments in Astrophysics Through Multi-Messenger Observations
of Gravitational Wave Sources'', No.~24103006 (H.N., T.T., T.N.)
and by the Grant-in-Aid from MEXT of Japan No.~15H02087 (T.T., T.N.).
We gratefully acknowledge all participants in 
``Gravitational Wave Physics and Astronomy Workshop (GWPAW) 2015,''
held June 17--20, 2015 in Osaka, Japan.
H.~N. would like to thank Y.~Nishino for useful suggestions.

%%%%%%%%%%%%%%%%%%%%%%%%%%%%%%%%%%%%%%%%
%\appendix

%%%%%%%%%%%%%%%%%%%%%%%%%%%%%%%%%%%%%%%%

%%%%%%%%%%%%%%%%%%%%%%%%%%%%%%%%%%%%%%%%
\end{document}